\documentclass[12pt,preprint]{aastex}

\bibpunct{(}{)}{;}{a}{}{,}

\begin{document}

   \title{The kinematics of a globally propagating disturbance in the solar corona}

   \author{David M. Long, Peter T. Gallagher, R. T. James McAteer, ~\and D. Shaun Bloomfield}

 \email{longda@tcd.ie}

   \affil{Astrophysics Research Group, School of Physics, Trinity College Dublin, Dublin 2, Ireland.}

  \begin{abstract}
  
The kinematics of a globally propagating disturbance (also known as an ``EIT wave") is discussed using Extreme UltraViolet Imager (EUVI) data from \textit{Solar Terrestrial Relations Observatory} (\textit{STEREO}). We show for the first time that an impulsively generated propagating disturbance has similar kinematics in all four EUVI passbands (304, 171, 195, and 284~\AA). In the 304~\AA\ passband the disturbance shows a velocity peak of 238$\pm$20~km~s$^{-1}$ within $\sim$28 minutes of its launch, varying in acceleration from 76~m~s$^{-2}$ to -102~m~s$^{-2}$. This passband contains a strong contribution from a \ion{Si}{11} line (303.32~\AA) with a peak formation temperature of $\sim$1.6~MK. The 304~\AA\  emission may therefore be coronal rather than chromospheric in origin. Comparable velocities and accelerations are found in the coronal 195~\AA\ passband, while lower values are found in the lower cadence 284~\AA\ passband. In the higher cadence 171~\AA\ passband the velocity varies significantly, peaking at 475$\pm$47~km~s$^{-1}$ within $\sim$20 minutes of launch, with a variation in acceleration from 816~m~s$^{-2}$ to -413~m~s$^{-2}$. The high image cadence of the 171~\AA\ passband (2.5 minutes compared to 10 minutes for the similar temperature response 195~\AA\ passband) is found to have a major effect on the measured velocity and acceleration of the pulse, which increase by factors of $\sim$2 and $\sim$10, respectively. This implies that previously measured values (e.g., using EIT) may have been underestimated. We also note that the disturbance shows strong reflection from a coronal hole in both the 171 and 195~\AA\ passbands. The observations are consistent with an impulsively generated fast-mode magnetoacoustic wave.

\end{abstract}

\keywords{Sun: corona -- Sun: flares}

\section{Introduction}

   Observations of globally propagating coronal disturbances at non-radio wavelengths were first achieved by the EUV Imaging Telescope (EIT) on board the \emph{Solar and Heliospheric Observatory (SOHO)} spacecraft \citep{Thompson:1998sf}. These ``EIT waves" were thought to be the coronal counterpart of the chromospheric Moreton wave, first seen by Moreton and associates \citep{Moreton:1960fj, Athay:1961yq}. Moreton waves were originally explained as the ``skirt'' of a coronal fast-mode wave caused by a flare \citep{Uchida:1968qv}. Upon the discovery of EIT waves it was assumed that they were the coronal waves described by Uchida, but differences between EIT and Moreton waves question this assumption. EIT waves are observed to be broad and diffuse, forming circular wave fronts when unimpeded, and can coherently travel the entire solar disk. In contrast, Moreton waves are observed to be strongly defined, narrow semi-circular fronts \citep{Warmuth:2004mb}. EIT waves and Moreton waves have been observed travelling co-spatially \citep{Thompson:2000bh, Okamoto:2004cq} and Moreton waves are always observed in conjunction with EIT waves. However, the inverse is not necessarily true \citep{Wills-Davey:2007oa}. 
      
     Propagating disturbances have been observed in the soft X-ray \citep[SXR;][]{Khan:2002qq}, the extreme ultraviolet \citep[EUV;][]{Thompson:1998sf}, radio \citep{White:2005ss}, H$\alpha$ ~\citep{Moreton:1960fj} and also in \ion{He}{1} 10830~\AA\ ~\citep{Gilbert:2004hl}. Any theory proposed must be able to explain these observations satisfactorily. At present there are two main competing theories.
     
      The first theory considers Moreton waves to be produced by the legs of a piston-driven shock (induced by a rising flux rope) which straddles an expanding coronal mass ejection (CME). The wave is observed as the legs of the shock sweep the chromosphere. In this theory, the ``EIT wave'' is formed by successive opening of the field lines covering the flux rope. The net result is that EIT waves are not real ``waves'', but are the propagation of perturbation sources due to a CME lift-off \citep{Delannee:1999mz, Chen:2002rw, Chen:2005xe}. This theory is consistent with statistical studies showing that ``EIT waves'' are more closely associated with CMEs than solar flares \citep{Biesecker:2002lq, Cliver:2005bd}. However, the theory relies on the simultaneous occurrence of Moreton and EIT waves, making it inconsistent with EIT wave only observations \citep{Wills-Davey:2007oa}.
    
      The second theory describes EIT waves as fast-mode magnetoacoustic waves, based on the physics outlined by \citet{Uchida:1968qv}. Fast magnetoacoustic waves can reproduce the brightenings observed in EUV images as well as explaining some of the characteristics of EIT waves, including their slow expansion rates \citep{Wang:2000tg}. This may also explain reflection and refraction of  EIT waves at coronal hole and active region boundaries \citep{Veronig:2006fy}. This interpretation has been used to create numerical simulations which closely imitate the observed data \citep{Wu:2001dz, Ofman:2002pi, Ofman:2007pr}. In addition, previous studies of the power spectra of EIT waves appear to confirm a periodic wavelike behaviour \citep{Ballai:2005kc}. The minimum fast-mode speed of a magnetoacoustic wave in a magnetised plasma is constrained by the Alfv\'{e}n speed of the plasma. However, previous EIT observations yield speeds below typical values for the Alfv\'{e}n speed of the solar corona.
            
      In this Letter we describe an event occurring on 2007 May 19. In \S~2 we detail our observations and methods of data analysis, while in \S~3 we outline our results, before discussing them and giving our conclusions in \S~4.
       
\section{Observations and Data Analysis}

   On 2007 May 19 a GOES Class B9 flare occurred in active region NOAA 10956. It began at 12:34 UT, peaked at 13:02 UT, and ended at 13:19 UT. According to the \emph{SOHO}/LASCO CME Catalog, an associated CME first appeared in the C2 field of view at 13:24:04 UT. In addition, a disappearing solar filament was observed erupting from NOAA 10956, beginning at 12:31 UT.
     
\begin{figure*}[!ht]
\centering
\plotone{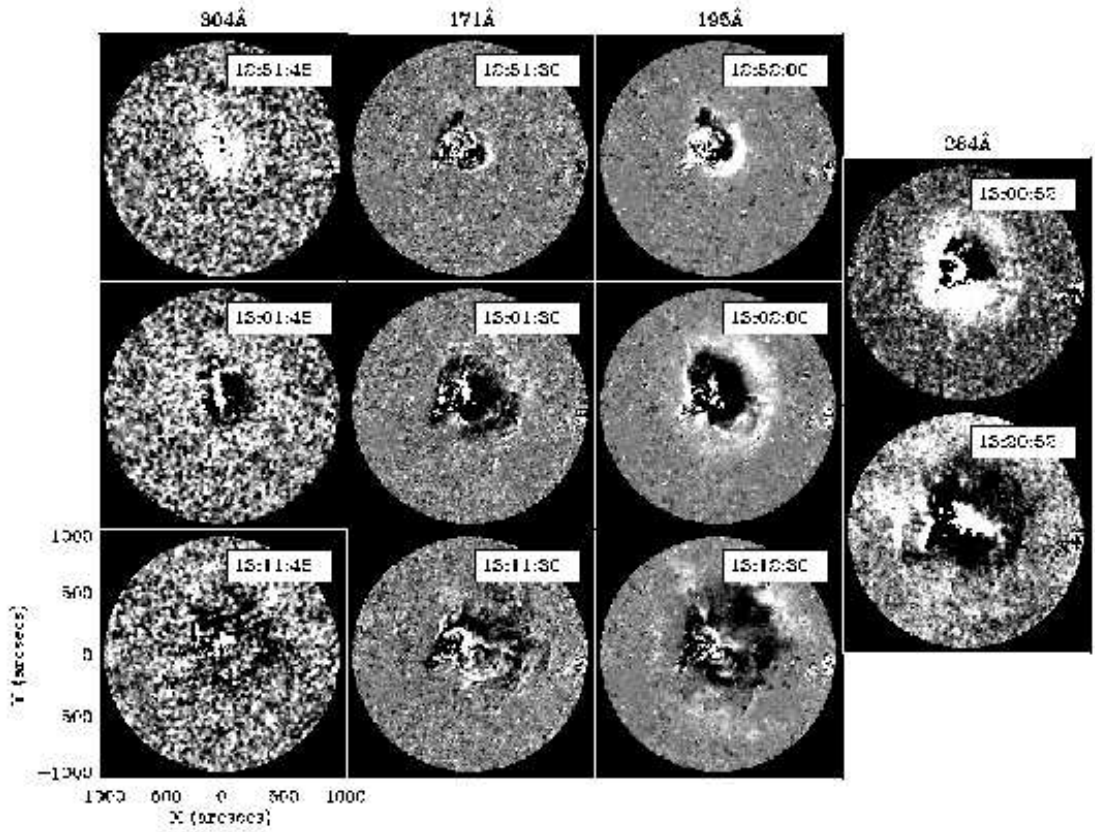}
\caption{Evolution of the disturbance with time in all four EUVI wavelengths from \textit{STEREO-A}. The wavelengths are arranged from left to right as 304~\AA, 171~\AA, 195~\AA\ and 284~\AA. Running difference movies are available online for each of the four passbands.}
\label{fig:images}
\end{figure*}

   The Extreme UltraViolet Imager (EUVI), part of the Solar Earth Connection Coronal and Heliospheric Investigation (SECCHI) suite of instruments onboard both \textit{STEREO-A} and \textit{STEREO-B}, was used for this work. Unfortunately, \textit{SOHO}/EIT was in CCD bakeout at the time and was unavailable. The cadence was  600~s for 304~\AA, 150~s for 171~\AA, 600~s for 195~\AA\ and 1200~s for 284~\AA. Each image has a pixel scale of 1.6\arcsec~\citep{Wuelser:2004bs} and the peak sensitivities for each passband are approximately 0.07~MK (304~\AA), 1~MK (171~\AA), 1.5~MK (195~\AA) and 2.25~MK (284~\AA). The data were flat-fielded, degridded and corrected for on-board image processing and spacecraft rotation. The images were differentially derotated to 12:32~UT to account for solar rotation. Running difference images were produced by subtracting the previous image in time from each image. This process was repeated for the data in each of the four passbands from both spacecraft.  A boxcar filter was applied to each image with the filter widths varying according to passband (15, 9, 3 and 7 pixels for 304, 171, 195, and 284~\AA, respectively), chosen to improve visibility of the brightening without losing detail due to excessive smoothing. The resulting series of running difference images in each wavelength showed the propagation of the disturbance with time. A selection of these images may be seen in Figure~\ref{fig:images}, showing the bright front after formation, once expanded, and before it becomes too diffuse to see. In addition, running difference movies showing the propagation of the disturbance may be found online.
        
\begin{figure}
\begin{center}
\plotone{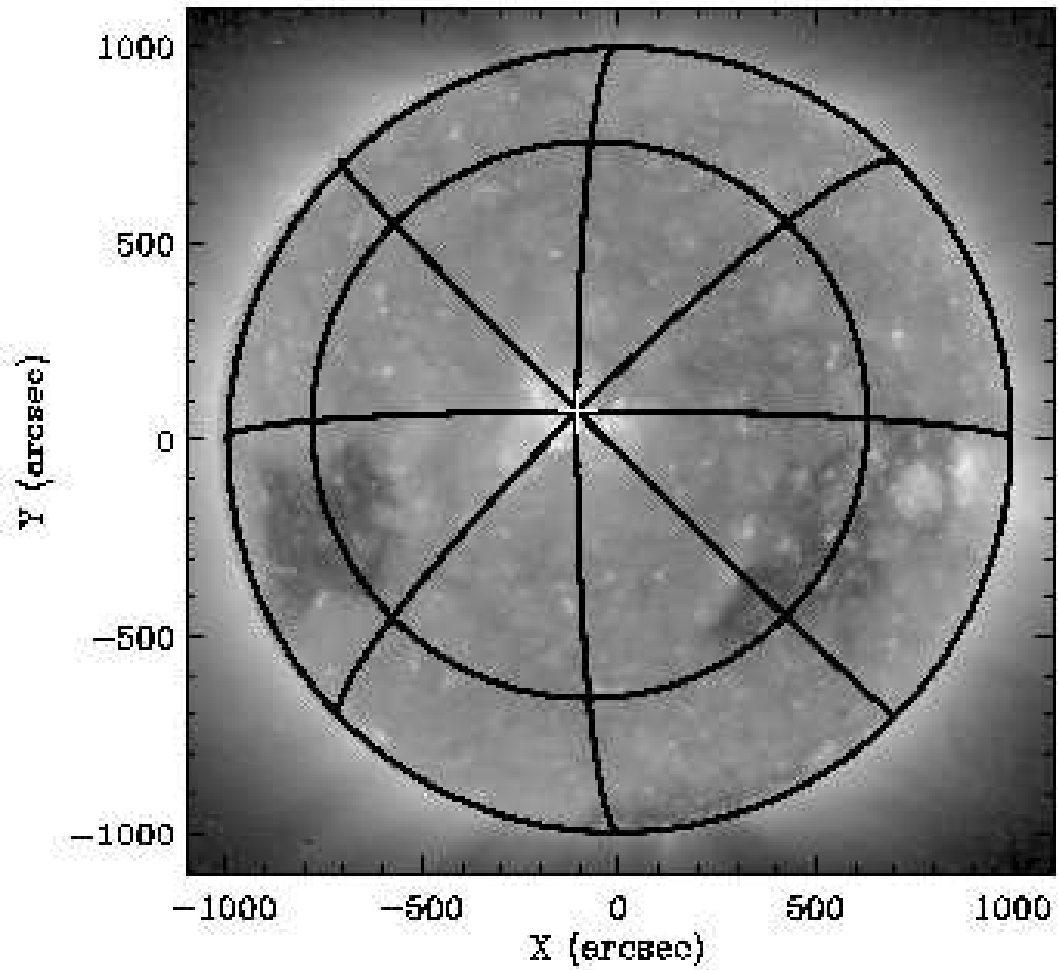}
\caption{\textit{STEREO-A} 195~\AA\ image with heliographic grid overlay. The north pole of the grid is centered on the flare kernel.}
\label{fig:grid}
\end{center}
\end{figure}

     A transformed heliographic grid was plotted over the difference images to account for the spherical nature of the Sun, with the north pole centered on the flare kernel. The transformed latitude and longitude grid lines are separated by angles of 45 degrees, as illustrated in Figure~\ref{fig:grid} for a 195~\AA\ image. The flare kernel was assumed to be the source of the disturbance. The grid lines were then used as a guide for point-and-click methods, allowing the temporal variation in the location of the front to be determined radially away from the source region along a chosen great circle. Distances were measured as angular separation from the flare kernel along a great circle, similar to \citet{Warmuth:2004mb} and \citet{Veronig:2006fy}. The grid was kept constant for all wavelengths from \textit{STEREO-A}, with a new grid chosen for \textit{STEREO-B} due to the different spacecraft locations. Measured angular separations were converted to distance on the assumed spherical surface, resulting in the upper panels of Figure~\ref{fig:graphs}. This allowed velocity and acceleration graphs to be calculated, as seen in the the middle and lower panels of Figure~\ref{fig:graphs}. 

\section{Results}

     The second row of difference images in Figure~\ref{fig:images} clearly shows a bright front on the solar disk in all four passbands. While "EIT waves'' have previously been observed in 195~\AA\, this is the first time it has been identified in the 304, 171 and 284~\AA\ passbands. The difficulty in directly comparing the results from observations of unequal cadence is highlighted by the velocity-time graphs in the middle row of Figure~\ref{fig:graphs}, which depicts each wavelength at its original cadence. The curve corresponding to the 171~\AA\ images shows a significant peak in velocity (increasing to 475$\pm$47~km~s$^{-1}$ within $\sim$20 minutes) and a significant variation in the acceleration (from 816~m~s$^{-2}$ to -413~m~s$^{-2}$). This is distinctly different from the more gradual velocity peaks observed in 195, 304, and 284~\AA\ and subsequent small variations in their acceleration curves: peak velocity of 262$\pm$4~km~s$^{-1}$ within $\sim$28 minutes and acceleration varying from 93~m~s$^{-2}$ to -109~m~s$^{-2}$ for 195~\AA; peak velocity of 238$\pm$20~km~s$^{-1}$ within $\sim$28 minutes and acceleration varying from 76~m~s$^{-2}$ to -102~m~s$^{-2}$ in 304~\AA; peak velocity of 153$\pm$5~km~s$^{-1}$ within $\sim$27 minutes and acceleration varying from 23~m~s$^{-2}$ to -32~m~s$^{-2}$ for 284~\AA. These results from \textit{STEREO-A} are given in Table~\ref{table1}. Similar velocity and acceleration curves are seen from both \textit{STEREO} spacecraft, with differences between \textit{STEREO-A} and \textit{STEREO-B} probably due to geometrical effects.  
     
     The errors in the distance-time graphs were given by the smoothing kernel used for each passband, and are smaller than the symbol size. Errors in velocity and acceleration were calculated using a 3-point centre difference, while the first and last values were calculated using 3-point forward and reverse differences, respectively. This resulted in the error bars for the first and last data points being larger than for the middle points.  
 
      Every fourth image from 171~\AA\ was chosen for running difference analysis in order to compare data from 171, 195, and 304~\AA\ at the same image cadence. The effective time of the 171~\AA\ running difference images was chosen to be as close as possible to that of the 195~\AA\ running difference images, yielding the results in Figure~\ref{fig:graphs2}. The data for 284~\AA\ has been removed for Figure~\ref{fig:graphs2} as it has a cadence of 20 minutes, and is therefore not comparable to the 304, 171, and 195~\AA\ data. The peak in velocity is more gradual than that for 171~\AA\ at a cadence of 2.5 minutes and the acceleration-time plot, while still showing acceleration followed by deceleration, is not as pronounced as in Figure~\ref{fig:graphs}. The 10 minute cadence 171~\AA\ distance, velocity and acceleration curves are practically identical to the 10 minute cadence 304 and 195~\AA\ curves.
      
\section{Conclusions}

     The coronal propagation event on 2007 May 19 was studied using images from \textit{STEREO}/EUVI. The disturbance was visible in all four passbands allowing measurement of the variation of the distance, velocity and acceleration of the front with time. The higher cadence of 171~\AA\ images allowed for a more detailed analysis than in the other three passbands. It is shown by degrading the cadence of the 171~\AA\ running difference images that the results obtained for 195 and 304~\AA\ are most likely due to temporal undersampling. We suggest that this temporal undersampling may explain the lower velocities previously measured in EIT data (12-minute cadence), which may have missed short time-scale variations. 
     
\begin{figure*}
\begin{center}
\begin{tabular}{cc}
\plottwo{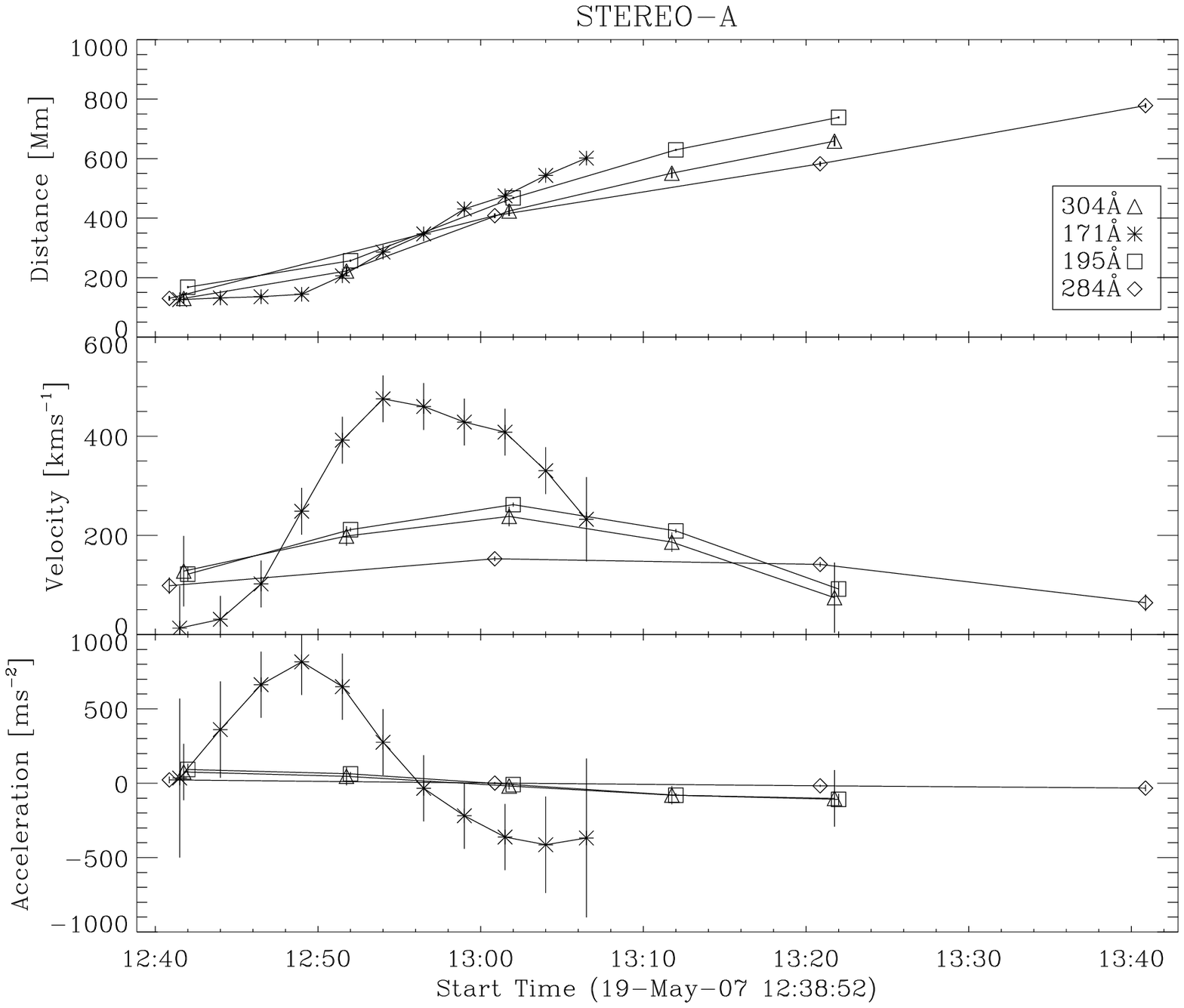}{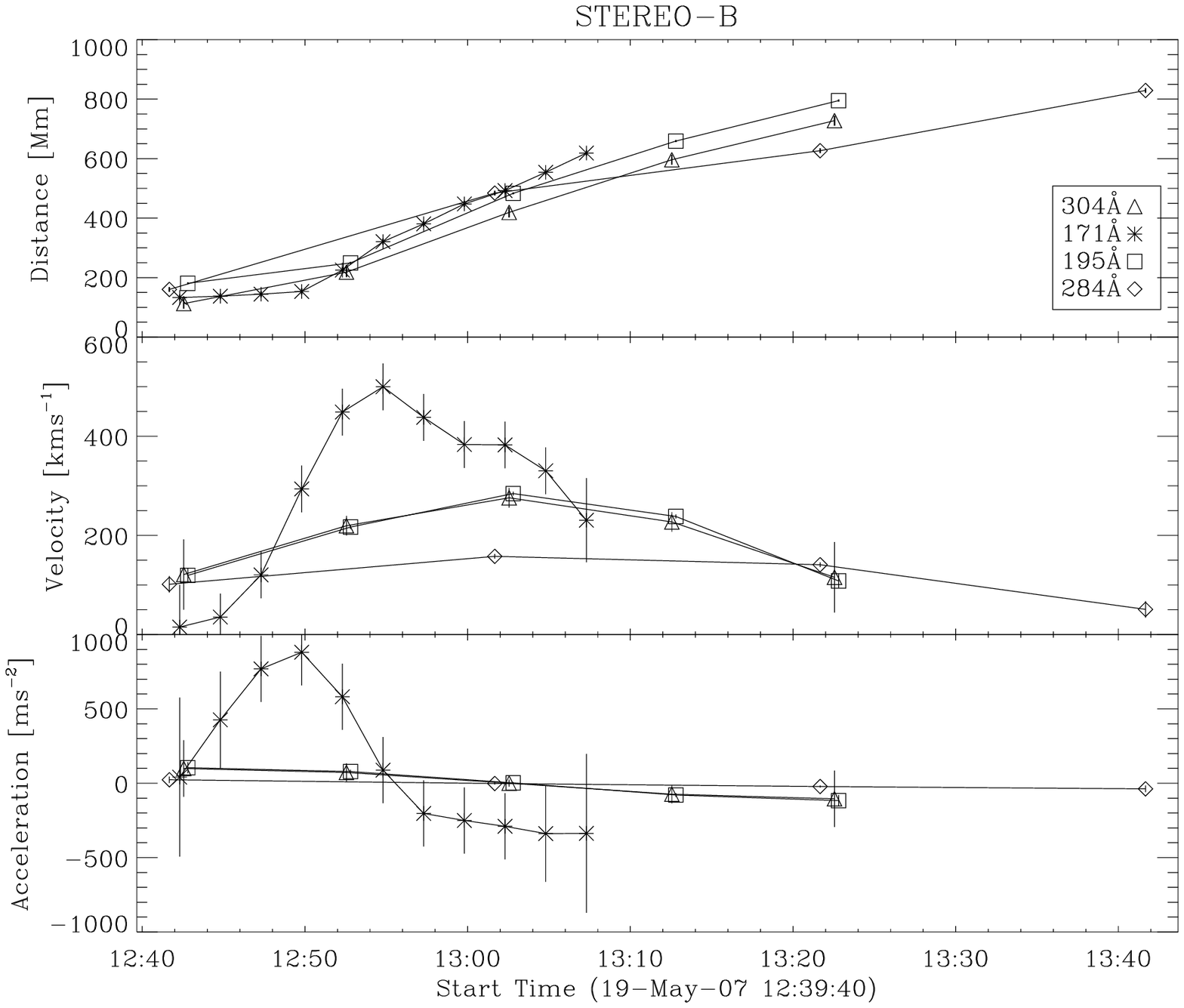}
\end{tabular}
\caption{Distance-time (top), velocity-time (middle), and acceleration-time (bottom) plots for \textit{STEREO-A} (left) and \textit{STEREO-B} (right). Distances are measured from the flare kernel to the leading edge of the running difference brightening along the great circle longitude line to solar west (see Fig.~\ref{fig:grid}). The first and last data point error bars in the 171~\AA\ data have been divided by two for display purposes.}
 \label{fig:graphs}
\end{center}
 \end{figure*}

\begin{figure}
\begin{tabular}{c}
\plotone{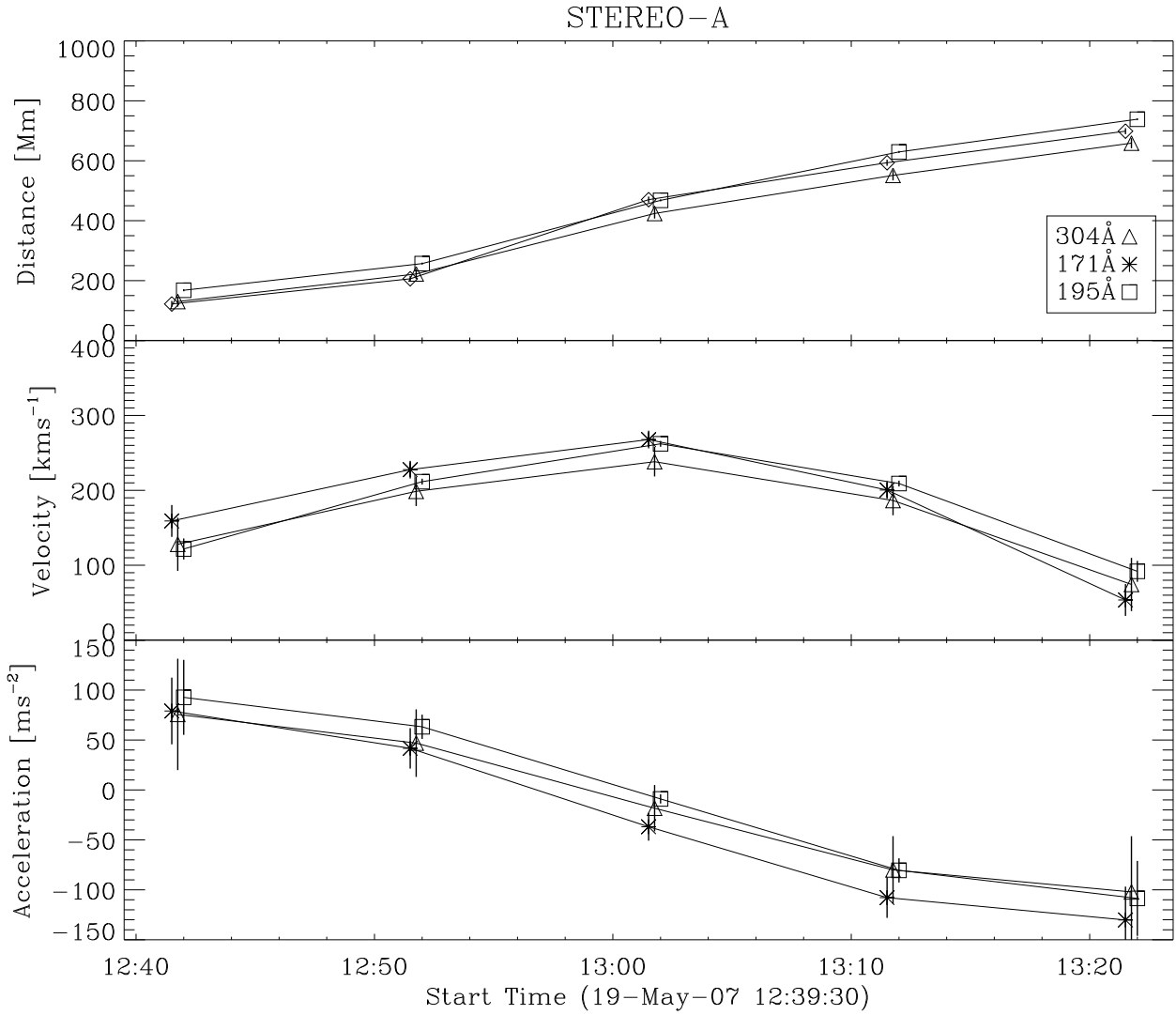}
\end{tabular}
\caption{Same as Fig.~\ref{fig:graphs}, but with 171~\AA\ at 10 minute cadence and excluding 284~\AA.}
\label{fig:graphs2}
\end{figure}

      In the 171~\AA\ images the front shows increasing positive acceleration to a maximum value of 816~m~s$^{-2}$ followed by a decrease to -413~m~s$^{-2}$ -- i.e., it accelerates away from the flare site before decelerating once further away. These data yield an average front velocity of 196~km~s$^{-1}$, with a peak at 475$\pm$47~km~s$^{-1}$ which is at the higher end of the previously measured range of ``EIT wave'' velocities. These results from \textit{STEREO-A} are qualitatively corroborated by \textit{STEREO-B} (c.f., Figure~\ref{fig:graphs}).  
        
      Although quite faint in the raw images, this event was visible in 304~\AA\ running difference images and is the first time that a coronal ``EIT wave'' has been observed in the 304~\AA\ passband.  The data required extensive filtering to make the disturbance visible above variable small--scale features present in the running difference images. It should be noted that there can be a strong contribution from the \ion{Si}{11}~303.32~\AA\ line in the 304~\AA\ passband, which has a peak formation temperature of $\sim$1.6~MK \citep{Brosius:1996qv}. It is therefore possible that the disturbance seen in the 304~\AA\ passband is coronal in origin rather than chromospheric. 
    
    The properties of this event should allow us to clarify whether it was a fast-mode magnetoacoustic wave or the propagation of perturbation sources due to CME lift-off. Assuming that the wave propagates at right angles to the predominantly radial magnetic field orientation in the quiet Sun, the wave will propagate as a fast-mode magnetoacoustic wave. The characteristic velocity can therefore be calculated using $v_{f}=(c_{s}^2 + v_{A}^2)^{1/2}$, where $c_{s}=(\gamma k_{B}T/m)^{1/2}$ is the sound speed and $v_{A} = B/(4\pi\rho)^{1/2}$ is the Alfv\'{e}n speed. Assuming typical coronal values for the plasma density and magnetic field, the fast-mode propagation speed is in the range 200--1000~km~s$^{-1}$. The propagations speeds found in this work are higher than previous ``EIT wave'' values and are now more consistent with the fast magnetoacoustic wave interpretation. 
    
	In addition, we note that the event was associated with a partial halo CME and a filament lift-off which began at $\sim$12:31~UT. The fact that the timescale involved for the front to accelerate to its peak velocity is comparable to that for a CME ($\sim$30 minutes) may indicate a common launch mechanism. However, the observed rise and fall in velocity is in contrast to the \citet{Delannee:2007uq} model which predicts a constantly increasing velocity. In addition, the disturbance was reflected from a coronal hole to the south--west of the launch site, as observed in the 171 and 195~\AA\ passbands. This provides further support for the coronal wave interpretation of "EIT waves''.
      
\acknowledgements
DML is supported by a Studentship from the Irish Research Council for Science, Engineering and Technology (IRCSET); DSB by an ESA/PRODEX award administered by Enterprise Ireland; and RTJMcA by a Marie Curie Intra-European Fellowship. We thank R. Erd\'{e}lyi and the referee for comments which improved this Letter, and the STEREO/SECCHI consortium for providing open access to their data and technical support.

\clearpage

\begin{deluxetable}{ccccc}
\tablecolumns{5}
\tabletypesize{\small}
\tablewidth{0pt}
\centering
\tablecaption{Velocities and accelerations from \textit{STEREO-A}.\label{table1}}
\tablehead{
\colhead{Passband} & \colhead{Cadence} & \colhead{$v_{max}$} & \colhead{$a_{max}$} & \colhead{$a_{min}$} \\ 
\colhead{(\AA)} & \colhead{(minutes)} & \colhead{(km~s$^{-1}$)} & \colhead{(m~s$^{-2}$)} & \colhead{(m~s$^{-2}$)}
}
\startdata
304 & 10 & 238 & 76 & -102 \\
171 & 2.5 & 475 & 816 & -413 \\
195 & 10 & 262 & 93 & -109 \\
284 & 20 & 153 & 23 & -32
\enddata
\tablecomments{Errors: $\leq$50~km~s$^{-1}$; $\leq$200~m~s$^{-2}$.}
\end{deluxetable}

\clearpage

\clearpage

\clearpage

\clearpage
 
\begin{thebibliography}{}
\bibitem[Athay \& Moreton(1961)]{Athay:1961yq}Athay, R. G. \& Moreton, G. E. 1961, \apj, 133, 935
\bibitem[Ballai et al.(2005)]{Ballai:2005kc}Ballai, I., Erd\'{e}lyi, R., Pint\'{e}r, B. 2005, \apj, 633, 145
\bibitem[Biesecker et al.(2002)]{Biesecker:2002lq}Biesecker, D. A., Myers, D. C., Thompson, B. J., Hammer, D. M., \& Vourlidas, A. 2002, \apj, 569, 1009
\bibitem[Brosius et al.(1996)]{Brosius:1996qv}Brosius, J. W., Davila, J. M., Thomas, R. J., Monsignori-Fossi, B. C. 1996, \apj, 106, 143
\bibitem[Chen et al.(2005)]{Chen:2005xe}Chen, P. F., Fang, C., \& Shibata, K. 2005, \apj, 622, 1202
\bibitem[Chen et al.(2002)]{Chen:2002rw}Chen, P. F., Wu, S. T., Shibata, K., \& Fang, C.2002, \apj, 572, L99
\bibitem[Cliver et al.(2005)]{Cliver:2005bd}Cliver, E. W., Laurenza, M., Storinin, M., \& Thompson, B.J. 2005, \apj, 631, 604
\bibitem[Delann\'{e}e \& Aulanier(1999)]{Delannee:1999mz}Delann\'{e}e, C. \& Aulanier, G. 1999, \solphys, 190, 107
\bibitem[Delann\'{e}e et al.(2007)]{Delannee:2007uq}Delann\'{e}e, C., T\"{o}r\"{o}k, T., Aulanier, G. \& Hochedez, J.-F. 2007, \solphys, 247, 123
\bibitem[Gilbert \& Holzer(2004)]{Gilbert:2004hl}Gilbert, H. R. \& Holzer, T. E. 2004, \apj, 610, 572
\bibitem[Khan \& Aurass(2002)]{Khan:2002qq}Khan, J. I. \& Aurass, H. 2002, \aap, 383, 1018
\bibitem[Moreton \& Ramsey(1960)]{Moreton:1960fj}Moreton, G. E. \& Ramsey, H. E. 1960, \pasp, 72, 357
\bibitem[Ofman(2007)]{Ofman:2007pr}Ofman, L. 2007, \apj, 655, 1134
\bibitem[Ofman \& Thompson(2002)]{Ofman:2002pi}Ofman, L. \& Thompson, B. J. 2002, \apj, 574, 440
\bibitem[Okamoto et al.(2004)]{Okamoto:2004cq}Okamoto, T. J., Nakai, H., Keiyama, A., Narukage, N., UeNo, S., Kitai, R., Kurokawa, H., \& Shibata, K. 2004, \apj, 608, 1124
\bibitem[Thompson et al.(1998)]{Thompson:1998sf}Thompson, B. J., Plunkett, S. P., Gurman, J. B., Newmark, J. S., St. Cyr, O. C., \& Michels, D. J. 1998, \grl, 25, 2465
\bibitem[Thompson et al.(2000)]{Thompson:2000bh}Thompson, B. J., Reynolds, B., Aurass, H., Gopalswamy, N., Gurman, J. B., Hudson, H. S., Martin, S. F., \& St. Cyr, O. C. 2000, \solphys, 193, 161
\bibitem[Uchida(1968)]{Uchida:1968qv}Uchida, Y. 1968, \solphys, 4, 30
\bibitem[Veronig et al.(2006)]{Veronig:2006fy}Veronig, A. M., Temmer, M., Vr\H{s}nak, B., \& Thalmann, J. K. 2006, \apj, 647, 1466
\bibitem[Wang(2000)]{Wang:2000tg}Wang, Y.-M. 2000, \apj, 543, L89
\bibitem[Warmuth et al.(2004a)]{Warmuth:2004mb}Warmuth, A., Vr\H{s}nak, B., Magdaleni\'{c}, J., Hanselmeier, A., Otruba, W. 2004, \aap, 418, 1101
\bibitem[White \& Thompson(2005)]{White:2005ss}White, S. M. \& Thompson, B. J. 2005, \apj, 620, L63
\bibitem[Wills-Davey et al.(2007)]{Wills-Davey:2007oa}Wills-Davey, M. J., DeForest, C. E., \& Stenflo, J. O. 2007, \apj, 664, 556
\bibitem[Wu et al.(2001)]{Wu:2001dz}Wu, S. T., Zheng, H., Wang, S., Thompson, B. J., Plunkett, S. P., Zhao, X. P., \& Dryer, M. 2001, \jgr, 106, 25089
\bibitem[Wuelser et al.(2004)]{Wuelser:2004bs}Wuelser, J.-P., Lemen, J. R., \& Tardell, T. D. 2004 in \procspie, 5171, ed. S. Fineschi, M. A. Gummin, 111
\end{thebibliography}
\end{document}